\documentstyle[psfig,here]{ptptex}
\def\ynu{y_{\nu}}
\def\ydu{y_{\triangle}}
\def\ynut{{y_{\nu}}^T}
\def\ynuv{y_{\nu}\frac{v}{\sqrt{2}}}
\def\ynuvt{{\ynut}\frac{v}{\sqrt{2}}}

\title{Neutrino Mixing in the Seesaw Model}

\author{Junya {\sc Hashida}
        \thanks{E-mail address :
                      jhashida@theo3.phys.sci.hiroshima-u.ac.jp}
        Takuya {\sc Morozumi}
        \thanks{E-mail address :
                      morozumi@theo.phys.sci.hiroshima-u.ac.jp}
        and Agus {\sc Purwanto}
        \thanks{E-mail address :
                      purwanto@theo3.phys.sci.hiroshima-u.ac.jp}
}
\inst{Department of Physics, Hiroshima University,
      Higashi-Hiroshima 739-8526, Japan}

\abst{
We study the neutrino mixing matrix (the MNS matrix) in the seesaw model.
By assuming a large mass hierarchy for the heavy right-handed
Majorana mass, we show that, in the diagonal Majorana base, 
the MNS matrix is determined by a unitary matrix, $S$, which transforms
the neutrino Yukawa matrix, $y_{\nu}$, into a triangular form, $y_{\triangle}$.
The mixing matrix of light leptons is 
$V_{KM} S^{\prime *}$ , 
where 
$V_{KM} \equiv {V_{Le}}^{\dagger} V_{L \nu}$ 
and $S^{\prime *} \equiv {V_{L\nu}}^{\dagger} S^*$.
Large mixing may occur 
without fine tuning of the matrix elements of $y_{\triangle}$
even if the usual KM-like matrix $V_{KM}$ is given by $V_{KM} =1$.   
This large mixing naturally may satisfy the
experimental lower bound of the mixing implied by
the atmospheric neutrino oscillation.
}

\begin{document}

\maketitle

\begin{picture}(5,2)
\put(320,250){HUPD-9912}
\end{picture}

\section{Introduction}

Since the observation of neutrino oscillations,\cite{SOL,ATM,SKC}
the mixing matrix of the lepton sector (the MNS matrix \cite{MNS}) has been
discussed in many works.
Among the various proposed explanations of the light
neutrino mass, the seesaw mechanism \cite{YAN} is an attractive scenario.
In this paper, we study the MNS matrix in the
framework of the seesaw model. Specifically,
we assume that the gauge group is $SU(2) \times U(1)$.
Since there are three light neutrinos, each one
may have its partner, a gauge singlet neutrino with
large Majorana mass.
Moreover, neutrino
oscillation experiments \cite{SOL,ATM,SKC}
suggest that there is mass hierarchy. \footnote{According
to recent analysis of the
solar and atmospheric neutrino data from Super-Kamiokande
experiments,\cite{SOL,ATM,SKC} the mass squared differences are
$\Delta m_{21}^2 \sim 10^{-5} [{\rm eV^2}]$ for the MSW solution,
$\Delta m_{21}^2 \sim 10^{-10} [{\rm eV^2}]$
for the vacuum solution, and
$\Delta m_{32}^2 \sim 10^{-3} [{\rm eV^2}]$. Imposing the conditions $m_2^2 \gg
m_1^2$ and $m_3 > m_2$,
the ratio
$\Delta m_{32}^2/\Delta m_{21}^2 \sim (m_3^2-m_2^2)/m_2^2$
is $10^{2}$ for the MSW solution
and $10^7$ for vacuum solution.
This implies $m_3^2 \gg m_2^2$
if the relation ${m_2}^2 \gg {m_1}^2$ is assumed. 
If this is the case, numerically, ${m_2 \over m_3}=10^{-1}$
for the MSW solution and ${m_2 \over m_3}=10^{-3.5}$ for vacuum
solution.
}
In the context of the seesaw model, the mass hierarchy may 
originate from the Yukawa term and/or Majorana mass term.
To account for the hierarchy and the
mixing, there are two extreme explanations. The first explanation is that the hierarchy
comes from the Majorana mass term. The other explanation is that
the heavy right-handed neutrinos have a degenerate mass
and the hierarchy comes from the Yukawa term.
In this paper, we are interested in the former case.
If the mass hierarchy of neutrinos comes from the hierarchy of the Majorana
mass term $(m_{N1} \gg m_{N2} \gg m_{N3})$,
the diagonalization of the seesaw matrix is carried out
in a very different way than that for charged fermions.
This is what we discuss in this paper.
The seesaw mass matrix is  a $2N_f \times 2N_f$
matrix ($N_f=3$ in our case) and has the simple structure
\begin{eqnarray}
M_{\nu}=\left( \begin{array}{cc}
0 & \ynuv \\
\ynuvt & m_N \\
\end{array}
\right).
\label{MNU}
\end{eqnarray}
Its $N_f \times N_f$ submatrix corresponding to the
Majorana mass term of light neutrinos is zero, and the
submatrix corresponding to the Majorana mass term of heavy neutrinos
can be a real diagonal matrix.
The origin of the flavor mixing is the Dirac-type Yukawa
term denoted by $\ynuv$.
Because the seesaw matrix is a symmetric matrix, it can be
diagonalized by a unitary matrix $V$ as $V^T M_{\nu}V$.\cite{BIL}
Despite the simple structure,
it is not possible to diagonalize
the matrix analytically, because we have to treat a
$6 \times 6 $ matrix.
We perform the approximate
diagonalization of the mass matrix and obtain the
parametrization of $V$.
This produces
the mixing matrix of neutrino sector. Combining it
with the mixing matrix of the charged lepton sector, we may
determine the flavor mixing in the lepton sector, namely, the MNS matrix.
The method we employ is very similar to the diagonalization of
the seesaw matrix for the quark mass in
the context of a left-right model.\cite{LR}
In the model, isosinglet quarks
with a large mass hierarchy play a role similar
to the right-handed heavy neutrino in the present case.
An approximate diagonalization procedure of the seesaw type matrix
is developed 
in Ref. 8). % \cite{MOR}
In this paper, we extend that method to the
seesaw model for the neutrino mass with the same gauge group and
higgs as in the standard model.

The paper is organized as follows.
In section 2,
by demonstrating the procedure of the diagonalization in the
basis in which the Majorana mass term is diagonalized,
we explain how the MNS matrix comes out by introducing a triangular matrix.
In section 3,
we study the special case that the
Yukawa matrices for charged leptons and for neutrinos
are simultaneously diagonalized through a biunitary transformation.
If this were the case for the two Yukawa matrices
of up- and down-type quarks,
the Kobayashi-Maskawa (KM) matrix would be trivial.
However, in the seesaw model for the
neutrino mass, large mixing may still occur.
We also discuss the phenomenological implications
of our analysis.
In section 4,
we study the large mixing situation in the
weak basis in which the Yukawa matrices of charged leptons and
neutrinos are diagonal. In this basis, we study the texture
of the Majorana mass matrix, which leads to the large mixing, and
by doing so, we can understand the origin of the large mixing
from a different angle.
Finally, our conclusions are presented in section 5.

\section{\bf The MNS matrix in the Seesaw model}

We start with the
mass terms for lepton sector.
Without loss of generality,
the Majorana mass
matrix $m_N$ is assumed to be a real diagonal matrix,
${m_N}_{ij}={m_N}_i \delta_{ij}$.\cite{BIL}
Writing $M_{\nu}$ in Eq. (\ref{MNU})
in the diagonal Majorana mass basis,
the mass terms are
\begin{eqnarray}
-{\cal L}_{lepton}
&=&
\overline{e_L^0} y_e \frac{v}{\sqrt{2}} e_R^0
\nonumber \\
&+&
(\overline{{\nu_L^0}}~~\overline{{N_R^0}^c})
\left(\begin{array}{cccccc} 0&0&0&&&\\
                            0&0&0&&y_{\nu} \frac{v}{\sqrt{2}}&\\
                            0&0&0&&&\\
                            &&&m_{N1}&&\\
                            &{y_{\nu}}^T \frac{v}{\sqrt{2}}&&&m_{N2}&\\
                            &&&&&m_{N3}\\
                            \end{array}\right)
{{\nu_L^0}^c \choose N_R^0} + h.c. 
\nonumber \\
\label{L1}
\end{eqnarray}
We focus on the case that
the Majorana mass term is much larger than
Dirac mass term: $m_N \gg \ynuv$.
We are interested in the case that the diagonal Majorana
masses satisfy the relations ${m_N}_i \gg {m_N}_j$ ($i<j$), and
the rank of the Yukawa matrices $y_e$ and $y_{\nu}$ is $N_f$.

The approximate diagonalization procedure of the seesaw-type matrix
goes as follows.
$M_{\nu}$ is diagonalized by a $2N_f \times 2 N_f $ unitary
matrix
as $V^T M_{\nu} V= {M_{\nu}}_{\rm diagonal}$.
Let us parametrize this unitary matrix as\cite{BRA}
\begin{eqnarray}
V=\left(\begin{array}{cc}
        S & K \\
        T & A \\
        \end{array}
        \right),
\end{eqnarray}
where $S, K, T$ and $A$ are $ N_f \times N_f $ submatrices.
The unitarity relation is written in terms of the submatrices:
\begin{eqnarray}
S S^{\dagger}+ K K^{\dagger}=1,
\\
T T^{\dagger}+A A^{\dagger}=1,
\label{TTAA}
\\
S T^{\dagger}+ K A^{\dagger}=0.
\label{STKA}
\end{eqnarray}
Then ${M_{\nu}}_{\rm diagonal}$ is given by
\begin{eqnarray}
&&{M_{\nu}}_{\rm diagonal}=V^T M_{\nu} V=
\nonumber\\
&&\left( \begin{array}{cc}
T^T \ynuvt S + S^T \ynuv T  + T^T m_N T & T^T \ynuvt K +  S^T \ynuv A +
T^T m_N A  \\
(T^T \ynuvt K +  S^T \ynuv A + T^T m_N A)^T  &A^T \ynuvt K +
K^T \ynuv A + A^T m_N A \\
\end{array} \right).
\nonumber\\
\end{eqnarray}
We can easily see that the  submatrices
\begin{eqnarray}
&&T^T \ynuvt S + S^T \ynuv T  + T^T m_N T, 
\label{sub1} \\
&&  A^T \ynuvt K + K^T \ynuv A
+ A^T m_N A
\label{sub2}
\end{eqnarray}
must be diagonal, and the off-diagonal submatrix becomes zero:
\begin{eqnarray}
T^T \ynuvt K +  S^T \ynuv A + T^T m_N A =0.
\label{sub3}
\end{eqnarray}
In order for the  $O(m_N)$ part of Eq. (\ref{sub2}) to be diagonal,
we must have $A=1$ , 
\footnote{
We can also choose another $A$, {\it i.e.}
$A=\left( 
\begin{array}{ccc}
0 & 0 & 1  \\
0 & 1 & 0  \\
1 & 0 & 0
\end{array}
\right)$.
This choice changes only $K$ as $K=-S T^{\dagger} A$, and 
leaves the Majorana mass term in a diagonal form, 
but with a reversed sequence of mass elements, 
{\it i.e.} ${\rm diag}(m_{N1},m_{N2},m_{N3}) 
\rightarrow {\rm diag}(m_{N3},m_{N2},m_{N1})$.
In our approximation neither $S$ nor $T$ changes.
This fact in turn implies a light neutrino submatrix [Eq. (\ref{sub1})] 
of the same form 
as in the diagonal $A$ case.
}    
because $m_N$ is diagonal and its diagonal elements
are not degenerate. Then Eq. (\ref{TTAA}) implies that the leading order of
$T$ is O($\frac{v}{m_N}$). Equation (\ref{STKA}) means $K=O(\frac{v}{m_N})S$.
Therefore, to leading order, $S^{\dagger} S=1$.
In order for the $O(v)$ contribution of the RHS of Eq. (\ref{sub3}) to be
canceled,
we can neglect the contribution of the first term.
Then the relation $T=-\frac{v}{\sqrt{2}m_N} \ynu^T S$
is obtained. $K$ is also determined through Eq. (\ref{STKA}).
The above consideration leads to the following submatrices:
\begin{eqnarray}
A=1-O\left(\frac{v}{{m_N}}\right),
\quad S {S}^{\dagger}=1-O\left(\frac{v^2}{{m_N}^2}\right), \quad
T=-\frac{v}{\sqrt{2}m_N} \ynu^T S,\quad
K={\ynu}^*\frac{v}{\sqrt{2}m_N}.
\nonumber \\
\end{eqnarray}
Below, we neglect the deviation from unity of $A$ and $S S^\dagger$.
Now, the seesaw mass matrix is written as
\begin{eqnarray}
V^T {M_{\nu}} V=\left( \begin{array}{cc}
-S^T {\ynu} \frac{v^2}{2 m_N}{\ynu}^T S &0 \\
0 & m_N+ O(v) \\
\end{array}
\right).
\end{eqnarray}
At this stage, the unitary matrix is parametrized as
\begin{eqnarray}
V=\left(\begin{array}{cc}
1 & {\ynu}^* \frac{v}{\sqrt{2}m_N} \\
-\frac{v}{\sqrt{2}m_N}\ynu^T & 1 \\ \end{array} \right)
\left(\begin{array}{cc}
S & 0 \\
0 & 1 \\
\end{array}\right).
\end{eqnarray}
Further,$S$ must be chosen so that the
$-S^T \ynu \frac{v^2}{2m_N} \ynu^T S$ is diagonal.
For any matrix $y$ of rank $N_f$,
we can find a
unitary matrix which transforms $y$ into a triangular form.\cite{MOR}
We choose such matrix as $S^T$:
\begin{eqnarray}
S^T y_{\nu}=\left( \begin{array}{ccc}
                   y_{1}  & 0      & 0   \\
                   y_{21} & y_{2}  & 0   \\
                   y_{31} & y_{32} & y_3 \\
                   \end{array} \right)
           \equiv \ydu.
\label{TRI}
\end{eqnarray}
Here the diagonal elements are real, and the off-diagonal elements
are complex in general.
In this form, it is noted that the lightest neutrino $\nu_e$ only
interacts with $N_1$, $\nu_{\mu}$ couples to both $N_1$ and $N_2$,
and $\nu_{\tau}$ couples to all heavy neutrinos $N_1,~N_2$ and $N_3$,
as shown in Fig. 1.
By choosing $S$ in this way,
the seesaw mass matrix is approximately diagonalized.
This is shown by writing submatrix
$-S^T {\ynu} \frac{1}{m_N} {\ynu}^T
S$:
\begin{eqnarray}
-S^T {\ynu} \frac{1}{m_N} {\ynu}^T S
=
-\left(\begin{array}{ccc}
\frac{{y_1}^2}{m_{N1}}&
\frac{y_1 y_{21}}{m_{N1}} & \frac{y_1 y_{31}}{m_{N1}}
\\
\frac{y_1 y_{21}}{m_{N1}}  & \frac{{y_2}^2}{m_{N2}}
+ \frac{{y_{21}}^2}{m_{N1}} &  \frac{y_{21} y_{31}}{m_{N1}}+
\frac{y_2 y_{32}}{m_{N2}}
  \\
\frac{y_1 y_{31}}{m_{N1}} &
\frac{y_{21} y_{31}}{m_{N1}}+
\frac{y_2 y_{32}}{m_{N2}}
& \frac{{y_3}^2}{m_{N3}}+
\frac{{y_{32}}^2}{m_{N2}}+\frac{{y_{31}}^2}{m_{N1}}
  \\
\end{array}\right).
\label{LNU}
\end{eqnarray}
Because $m_{N1} \gg m_{N2} \gg m_{N3}$,
the off-diagonal elements are much smaller than the difference of
the diagonal elements.
Therefore the mass matrix is regarded as
approximately diagonal:
\begin{eqnarray}
V^T M_{\nu} V \simeq
\left(\begin{array}{cccccc} -\frac{{y_1}^2 v^2}{2 m_{N1}}&&&&&\\
                            &-\frac{{y_2}^2 v^2}{2 m_{N2}}&&&&\\
                            &&-\frac{{y_3}^2 v^2}{2 m_{N3}}&&&\\
                            &&&m_{N1}&&\\
                            &&&&m_{N2}&\\
                            &&&&&m_{N3}\\
                            \end{array}\right).
\end{eqnarray}
The above matrices show that
the mass of $\nu_e$ is
determined by the mass of the heaviest
singlet neutrino, $N_1$, and is not affected by the presence of the
other lighter singlet neutrinos, because $\nu_e$ couples to only $N_1$.
Moreover, though
$\nu_{\mu}$ couples to both
$N_1$ and $N_2$, its mass is mainly determined by the mass of $N_2$,
because the mass of $N_1$ is much larger than that of $N_2$.
Similarly, the mass of the $N_3$ mainly depends on the mass of $N_3$,
due to the relation $m_{N1} \gg m_{N2} \gg m_{N3}$.
If such a mass hierarchy of the three heavy neutrinos is
admitted, this procedure is a powerful method
for diagonalizing the seesaw-type mass matrix.

Then, the unitary matrix $V$ can be rewritten as
\begin{eqnarray}
V=
\left(\begin{array}{cc}
S & 0 \\
0 & 1 \\
\end{array}\right)
\left(\begin{array}{cc}
1 & {\ydu}^* \frac{v}{\sqrt{2} m_N} \\
-\frac{v}{\sqrt{2} m_N} {\ydu}^T & 1 \\
\end{array} \right).
\end{eqnarray}
The MNS matrix is a $3 \times 6$ submatrix of $V$ multiplied by
a $3 \times 3$ unitary matrix for diagonalization of
the Yukawa term of the charged lepton,
\begin{eqnarray}
V_{MNS}=({V_{Le}}^{\dagger} S^*) (1,\ydu\frac{v}{\sqrt{2}m_N}).
\end{eqnarray}
Here $V_{Le}$ comes from the diagonalization
of the Yukawa term of the charged lepton and is defined
by the equations
\begin{eqnarray}
&&e^0_L=V_{Le} e_L,\\
&&e^0_R=V_{Re} e_R,\\
&&y_{e D}={V_{Le}}^{\dagger} y_{e}{V_{Re}},
\end{eqnarray}
where $y_{eD}$ is a real diagonal matrix.
We introduce a unitary matrix which transforms the
Yukawa term of neutrinos $y_{\nu}$ to the diagonal form $y_{\nu D}$:
\begin{eqnarray}
y_{\nu}&=&V_{L\nu}y_{\nu D}{V_{R\nu}}^{\dagger}\nonumber  \\
       &=&S^* y_{\triangle}.
\label{Nu}
\end{eqnarray}
The second line of Eq. (\ref{Nu}) comes from Eq. (\ref{TRI}).
Therefore we can relate $y_{\triangle}$ and $y_{\nu D}$ through
the equation
\begin{eqnarray}
y_{\nu D}&=&{V_{L\nu}}^{\dagger} S^* y_{\triangle}V_{R\nu}
\nonumber \\
&=&S^{\prime *} y_{\triangle}V_{R\nu},
\label{YNU}
\end{eqnarray}
where we introduce another unitary matrix $S^{\prime}$,
\begin{equation}
{S^{\prime}}^*={V_{L\nu}}^{\dagger} S^*.
\end{equation}
$S^{\prime}$ represents the difference between the unitary matrix
which transforms $y_{\nu}$ into the triangular
form and the unitary matrix for the diagonalization
by a biunitary transformation.

Finally, we can write the
leptonic charged current as
\begin{eqnarray}
J_{L \mu}
&=&
( \bar{e}_L \bar{\mu}_L \bar{\tau}_L)
\gamma_{\mu} V_{MNS}
\left(\begin{array}{c}
{\nu_{L1}}\\
{\nu_{L2}}\\
{\nu_{L3}}\\
{N_{R1}}^c\\
{N_{R2}}^c\\
{N_{R3}}^c\\
\end{array} \right),
\label{CUR}
\end{eqnarray}
with
\begin{eqnarray}
V_{MNS}&=&({V_{Le}}^{\dagger} S^{*})
(1, \ydu \frac{v}{\sqrt{2} m_N}) 
\nonumber \\
&=&(V_{KM} S^{\prime *})
(1, \ydu \frac{v}{\sqrt{2} m_N}),
\end{eqnarray}
and
\begin{equation}
V_{KM}={V_{L e}}^{\dagger} V_{L \nu}.
\end{equation}
Here $\nu_L$ and ${N_R}^c$ are mass eigenstates of neutrinos.
From the first definition of $V_{MNS}$ from Eq. (\ref{CUR}), we can see that
the mixing of the neutrino sector is determined
by the unitary matrix which transforms the
neutrino Yukawa term into the triangular form.
In the second definition of  $V_{MNS}$ from Eq. (\ref{CUR}), we introduce
the usual KM-like matrix and rewrite the
matrix
${V_{Le}}^{\dagger} S^*$ as
$({V_{Le}}^{\dagger} V_{L\nu})
({V_{L\nu}}^{\dagger} S^*)=V_{KM} S^{\prime *}$.
By rewriting the matrix in this way,
we can divide the lepton mixing angles into
two parts,  one which comes from the usual
KM-like part, denoted by $V_{KM}$,
and another ($S^{\prime}$)
which comes from the difference between the unitary matrix $V_{L\nu}$
for biunitary transformation and the unitary matrix
$S$  for transformation into the triangular form.
The advantage of the latter decomposition is that both
$V_{KM}$ and $S^{\prime}$ are independent of the
choice of the weak basis.

We now summarize the results obtained in this section.
The MNS matrix is a product of the mixing matrix of the charged
lepton sector and that of the neutrino sector.
We find that the unitary matrix denoted by ${V_{Le}}^{\dagger} S^*$
determines the mixing matrix among light leptons.
Here $S$ denotes the unitary matrix which transforms the
neutrino Yukawa term into a triangular matrix.
$V_{Le}$ is the unitary matrix for the
diagonalization of the charged lepton Yukawa term.
This is proved in the
weak basis in which Majorana mass matrix of heavy neutrinos
is real diagonal and also under the assumption that
the heavy Majorana neutrinos have
greatly differing mass scales.

\section{Large mixing}

The MNS matrix obtained in the previous section is divided into
two parts unitary matrices $S^{\prime}$ and $V_{KM}$.
We note that $S^{\prime}$ is determined by
$y_{\nu}$ and $m_N$ in the diagonal Majorana basis,
while $V_{KM}$ is determined
by $y_{\nu} {y_{\nu}}^{\dagger}$ and $y_{e}{y_{e}}^{\dagger}$.
Therefore $S^{\prime}$ is determined by the mass
matrix of the neutrino sector, while $V_{KM}$ is determined
by the Yukawa terms only. In order for $V_{KM}$ to be a non-trivial
matrix, the following condition must be satisfied:
\begin{eqnarray}
V_{KM} \neq 1
\quad \leftrightarrow \quad
[y_{e}{y_{e}}^{\dagger}, y_{\nu}{y_{\nu}}^{\dagger}] \neq 0.
\end{eqnarray}
On the other hand, the condition for $S^{\prime}$ being non-trivial is
\begin{eqnarray}
S^{\prime} \neq 1
\quad \leftrightarrow \quad
(y_{\triangle} {y_{\triangle}}^{\dagger})_{ij} \neq 0
\quad {\rm for} \quad (i \neq j),
\label{LR2}
\end{eqnarray}
in the diagonal Majorana basis.
Equation (\ref{LR2}) comes from Eq. (\ref{YNU}), because
\begin{equation}
{S^{\prime}}^* {y_{\triangle}} {y_{\triangle}}^{\dagger} {S^{\prime}}^T
={y_{\nu D}}^2.
\label{Spr}
\end{equation}
The mixing angle is the product of $S^{\prime}$ and $V_{KM}$.
We note that for a given $y_{\triangle}$
we may determine $S^{\prime}$. Equation (\ref{LR2}) and (\ref{Spr})
imply that the off-diagonal elements in the triangular matrix
$y_{\triangle}$ induce the mixing in $S^{\prime}$.
If the off-diagonal elements of $y_{\triangle}$ vanish,
$S^{\prime}=1$.
However, since $V_{KM}$ is arbitrary,
in general, lepton mixing angles are also arbitrary.
Even if there is large mixing in $S^{\prime}$,
this can be canceled by $V_{KM}$. There may also be 
the case that $S^{\prime}$ is the unit matrix and
there is large mixing in $V_{KM}$.
Here we focus on the case that $V_{KM}=1$.
Equivalently, we impose the commutation relation
\begin{eqnarray}
[y_e {y_e}^{\dagger}, y_{\nu} {y_{\nu}}^{\dagger}]=0.
\label{COM}
\end{eqnarray}
If this is
the case, $S^{\prime}$
determines the mixing angles.

Now we show that large mixing occurs without fine tuning
of the matrix elements of $\ydu$. We illustrate this
by considering the two flavor case $(N_f=2)$ and  apply Eq. (\ref{COM}) to
the explanation of the large mixing of atmospheric neutrinos.
In the two flavor case,
$\ydu \ydu^{\dagger}$, $S^{\prime}$ and the approximately diagonalized
light neutrino mass matrix $m_\nu$
are respectively expressed as
\begin{eqnarray}
\ydu \ydu^{\dag}
=
\left(\begin{array}{cc}
   {y}_2^2                 &
   {y}_2 {y}_{32}^*     \\
   {y}_2 {y}_{32}       &
   |{y}_{32}|^2+{y}_3^2
\end{array}\right),
\end{eqnarray}
\begin{eqnarray}
S^{\prime}=\left(\begin{array}{cc}
   \cos \theta            & -\sin \theta e^{i \phi} \\
   \sin \theta e^{-i \phi} & \cos \theta
\end{array}\right),
\end{eqnarray}
with
\begin{eqnarray}
\tan 2 \theta
=\frac{2 | {y}_{32}| {y}_2}
{-{y}_2^2+|{y}_{32}|^2+{y}_3^2}
, \quad \phi=\arg({y}_{32}),
\label{TAN}
\end{eqnarray}
and
\begin{eqnarray}
m_{\nu}=-\left(\begin{array}{cc}
\frac{{y_2}^2}{m_{N2}}&
\frac{y_2 y_{32}}{m_{N2}}
\\
\frac{y_2 y_{32}}{m_{N2}}  & \frac{{y_3}^2}{m_{N3}}
+ \frac{{y_{32}}^2}{m_{N2}}
  \\
\end{array}\right)\frac{v^2}{2},
\label{numass}
\end{eqnarray}
where $m_{N2} \gg m_{N3}$ is assumed.
In order that  Eq. (\ref{numass}) can be regarded as an approximately
diagonal matrix, the sufficient conditions on the elements of
the Yukawa term $\ydu$ are
\begin{eqnarray}
{\cal O}(y_2) = {\cal O}(y_3) \equiv y,
\label{PA}
\end{eqnarray}
\begin{eqnarray}
{\cal O} \left( \frac{|y_{32}|}{y} \right) \le 1.
\label{PB}
\end{eqnarray}
The first condition is that the hierarchy of the light neutrino
masses comes from the hierarchy of the Majorana mass term,
$m_{N2} \gg m_{N3}$.
The second condition suppresses further mixing from the form of
Eq. (\ref{numass}).

In Fig. 2, we show $\sin^2 2\theta$ as a function of
$\frac{|y_{32}|}{y_2}$ by changing the ratio $r={\frac{y_3}{y_2}}$.
For $r=1$,
maximal mixing occurs at infinitesimally small $|y_{32}|=\epsilon$,
and the mixing decreases as $|y_{32}|$ increases.
For $r<1$, maximal mixing occurs at non-zero
$y_{32}$ because the diagonal elements are degenerate
for $|y_{32}| \neq 0$.
For $r>1$, the mixing angle is suppressed because
the diagonal elements cannot be degenerate.
For $r \neq 1$, small mixing is also possible for
$|y_{32}| \ll |y_3-y_2|$.
In Fig. 3, we constrain the ratios of the Yukawa couplings
by using the result obtained by the Super-Kamiokande, Eq. (\ref{SIN}).
The allowed range is between two curves
in the parameter space,
$(r, \frac{|y_{32}|}{y_2})$.
We can see that the experimental constraint can be easily satisfied
for the range of the parameters in Eqs. (\ref{PA}) and (\ref{PB}).
The curves are obtained with the equation
\begin{eqnarray}
\biggr(\frac{{y}_3}{{y}_2}\biggr)^2+
\biggr(\frac{|{y}_{32}|}{{y}_2}
-\frac{1}{\tan2\theta}\biggr)^2
=1+\frac{1}{\tan^2 2\theta},
\end{eqnarray}
where $\sin^2 2\theta$ takes the minimum and the maximum values
of the experimental constraint, Eq. (\ref{SIN}).
The qualitative features of Figs. 2 and 3 can be understood
by using the approximate formulae for $\tan 2\theta$ under the
conditions set by Eq. (\ref{PA}) and Eq. (\ref{PB}).
From Eq. (\ref{TAN}) we have
\begin{eqnarray}
\tan 2 \theta
=
\left\{\displaystyle\begin{array}{lc}
\frac{2 y}{|y_{32}|} \geq 2 &
   \mbox{for} \quad
|y_{32}|^2 > y_3^2 - y_2^2, \\
\frac{|{y}_{32}|}{y_3-y_2} &
   \mbox{for} \quad
|y_{32}|^2 < y_3^2 - y_2^2.
\end{array}\right.
\label{TAT}
\end{eqnarray}
The first line corresponds to the case that the diagonal elements
are very degenerate and the second line to the case that
they are not degenerate.
The first case of Eq. (\ref{TAT}) leads to
\begin{eqnarray}
\sin^2 2\theta \geq 0.8.
\end{eqnarray}
This can be compared with the result for atmospheric neutrinos.
The Super-Kamiokande collaboration \cite{SKC} has reported the range
\begin{eqnarray}
\sin^2 2\theta_{\rm atm} \geq 0.8.
\label{SIN}
\end{eqnarray}
Moreover, the second line in Eq. (\ref{TAT}) means that large mixing
occurs for $|y_{32}| \sim |y_3 - y_2|$.

Finally, we comment on the mass scale $m_{N3}$ of the lightest gauge
singlet neutrino. According to the Super-Kamiokande result,
the mass-squared difference of the $\mu$ and $\tau$ neutrinos is of order 
$10^{-3} {\rm [eV^2]}$. Assuming that this corresponds to the
mass squared of the $\tau$ neutrino and that the Yukawa coupling is the
same order of magnitude as the gauge coupling, we obtain
\begin{eqnarray}
m_{\nu \tau} \simeq \frac{{M_W}^2}{m_{N3}}.
\end{eqnarray}
Therefore, the mass of the lightest gauge singlet neutrino
may be ${\cal O}(10^{14})$ [GeV].\\

\section{Texture of the Majorana mass}

In the previous section, we obtained large mixing
under the additional constraint that
the Yukawa matrices of charged leptons, $y_e y_e^{\dagger}$, and
neutrinos, $y_{\nu} y_{\nu}^{\dagger}$, 
are simultaneously diagonalized, 
{\it i.e.} $V_{Le}=V_{L\nu}$ [see Eq. (\ref{COM})].
The origin of the mixing may be traced to the presence of the
off-diagonal element of the triangular matrix in the diagonal
Majorana mass basis.
It is useful to study the same situation from a different viewpoint.
In this section, we study the large mixing case in
a different basis.
We can transform from the
diagonal Majorana mass basis to the diagonal basis of Yukawa terms
for charged leptons and neutrinos. If
we further impose the condition Eq. (\ref{COM}),
the transformation becomes a transformation of the
weak basis. Therefore $V_{MNS}$ becomes the $3 \times 3$
unit matrix.
In the new basis, the Majorana mass term is not diagonal, and
we may investigate the texture of the Majorana matrix that
induces large mixing.

In this basis, the mass term (\ref{L1}) is rewritten as
\begin{eqnarray}
-{\cal L}_{lepton}
=
\overline{e}_L {y_{eD}} \frac{v}{\sqrt{2}} e_R
+(\overline{\nu^{\prime}_L}~~\overline{{N^{\prime}_R}^c})
\left(\begin{array}{cc} 0 &
y_{\nu D} \frac{v}{\sqrt{2}}\\
y_{\nu D} \frac{v}{\sqrt{2}}& {V_{R \nu}}^T m_N V_{R \nu}
\end{array}\right)
{{\nu^{\prime}_L}^c \choose N^{\prime}_R} + h.c.,
\nonumber \\
\label{L2}
\end{eqnarray}
where
\begin{eqnarray}
&&\nu^{\prime}_L={V_{L\nu}}^{\dagger} \nu_L^0, \\
&& N^{\prime}_R={V_{R \nu}}^{\dagger} N^0_R,
\end{eqnarray}
and the leptonic charged current is
\begin{eqnarray}
J_{L \mu}
&=&
( \bar{e}_L \bar{\mu}_L \bar{\tau}_L)
V_{KM}
\gamma_{\mu}\left(\begin{array}{c}
{\nu_{L1}^{\prime}}\\
{\nu_{L2}^{\prime}}\\
{\nu_{L3}^{\prime}}
\end{array} \right).
\end{eqnarray}
Up to this point, we have not imposed the condition Eq. (\ref{COM}).
If this condition is imposed, namely $V_{KM}=1$,
the charged current becomes
\begin{eqnarray}
J_{L \mu}
=
( \bar{e}_L \bar{\mu}_L \bar{\tau}_L)
\gamma_{\mu}\left(\begin{array}{c}
{\nu_{L1}^{\prime}}\\
{\nu_{L2}^{\prime}}\\
{\nu_{L3}^{\prime}}
\end{array} \right).
\end{eqnarray}
${V_{R \nu}}^T m_N V_{R \nu}$ $(\equiv \tilde{m}_N)$
is the Majorana mass matrix in the basis of
the diagonal Yukawa term.
Because of Eq. (\ref{YNU}), $V_{R\nu}$ is determined by
${y_{\triangle}}^{\dagger} y_{\triangle}$ as
\begin{eqnarray}
{y_{\nu D}}^2=
{V_{R\nu}}^{\dagger}{y_{\triangle}}^{\dagger} y_{\triangle} V_{R\nu}.
\end{eqnarray}

Now we give the form of the Majorana mass matrix
that induces large mixing in the two flavor case $(N_f=2)$.
Because ${y_{\triangle}}^{\dagger} y_{\triangle}$ is expressed as
\begin{eqnarray}
{y_{\triangle}}^{\dagger} y_{\triangle}
=\left(\begin{array}{cc}
{y_2}^2+|y_{32}|^2 & {y_{32}}^* y_3 \\
y_{32} y_3         & {y_3}^2
\end{array}\right),
\end{eqnarray}
we can parametrize $V_{R \nu}$ as
\begin{eqnarray}
V_{R \nu}
=
\left(\begin{array}{cc}
\cos\alpha & -\sin\alpha~e^{i \beta} \\
\sin\alpha~e^{-i\beta} & \cos\alpha
\end{array}\right),
\end{eqnarray}
with
\begin{eqnarray}
\tan 2 \alpha =
\frac{2 |y_{32}| y_3}{{y_2}^2+|y_{32}|^2-{y_3}^2},
\quad \beta=-{\rm arg}(y_{32}).
\end{eqnarray}
Therefore, the Majorana mass matrix is written as
\begin{eqnarray}
\tilde{m}_N
= \left(\begin{array}{cc}
m_{N2} \cos^2 \alpha + m_{N3} \sin^2 \alpha~e^{-2i\beta} &
-\frac{1}{2}(m_{N2}e^{i\beta}-m_{N3}e^{-i\beta})\sin 2 \alpha \\
-\frac{1}{2}(m_{N2}e^{i\beta}-m_{N3}e^{-i\beta})\sin 2 \alpha &
m_{N2}\sin^2 \alpha~e^{2i\beta}+m_{N3}\cos^2 \alpha
\end{array}\right).
\nonumber\\
\end{eqnarray}

In Fig. 4, we display the $\sin^2 2\alpha$ as a function
$\frac{|y_{32}|}{y_2}$ by changing the ratio $r={\frac{y_3}{y_2}}$.
By comparing Figs. 3 and 4, we can see that
maximal mixing of the mixing angle $\theta$
and maximal mixing of $\alpha$ take place
at the same time only when the diagonal elements
of $y_{\triangle}$ are exactly degenerate ($r=1$).
In this case, by setting
$\sin^2 2 \alpha \sim 1$,
the Majorana mass matrix can be expressed as
\begin{eqnarray}
\tilde{m}_N
\simeq
\frac{m_{N2}}{2}\left(\begin{array}{cc}
1+ \epsilon e^{-2 i \beta} & -(1-\epsilon e^{-2i\beta})e^{i \beta} \\
-(1-\epsilon e^{-2i\beta})e^{i \beta}  & (1+\epsilon
e^{-2i\beta})e^{2i\beta}
\end{array}\right),
\label{Majo}
\end{eqnarray}
where $\epsilon=m_{N3}/m_{N2}$ $( \ll 1)$.
From Eq. (\ref{Majo}),
the texture which leads to large mixing has the 
characteristic feature that the
off-diagonal and diagonal elements are the same order of 
magnitude and the Majorana mass matrix is close to a democratic-type matrix.
There are small corrections to the democratic-type matrix,
which lead to the non-zero mass of the heavy neutrino
$N_3$.

\section{Conclusions and discussion}

We have studied the MNS matrix in the seesaw model.
We proposed an approximate procedure for
diagonalization of the seesaw matrix.
If we start with a real diagonal Majorana mass matrix
and assume a hierarchy for the masses of the heavy neutrinos,
we find that the unitary matrix $S$ that transforms
the Yukawa term into a triangular form determines the
mass eigenstate of light neutrinos up to a small mixing
of the heavy neutrinos. The triangular form has to be
arranged in such a way that the lighter $SU(2)$ doublet neutrinos couple
to the heavier singlet neutrinos. 
The mixing matrix for light leptons
is $V_{Le}^{\dagger} S^*$, {\it i.e.}, the product of the
unitary matrix $V_{Le}$ required by diagonalization of the Yukawa matrices
of charged leptons, $y_e y_e^{\dagger}$, and 
the unitary matrix $S^*$ to transform the Yukawa matrix of neutrinos, $y_{\nu}$,
into the triangular form $y_{\triangle}$.
By inserting the unitary matrix for the diagonalization of
the Yukawa matrices of the neutrinos $y_{\nu} y_{\nu}^{\dagger}$, 
the mixing matrix can be rewritten
as $V_{KM} {S^{\prime}}^*$. $V_{KM}$ is a usual KM-like matrix and
denotes the difference between the unitary matrices for the diagonalization
of Yukawa terms, and $S^{\prime}$ denotes the difference between the
unitary matrix for the transformation into the triangular form
and the unitary matrix for the diagonalization of $y_{\nu}$.
Therefore $S^{\prime}$ can be determined by the structure of
the mass matrix of the neutrino sector only. We find that
for a given $y_{\triangle}$, we can determine $S^{\prime}$.
We also find that the presence of the off-diagonal
element of $y_{\triangle}$ induces the mixing in $S^{\prime}$.
A large mixing angle is possible and it was explicitly
given for the two-flavor case.
Since $V_{KM}$ is arbitrary, we impose the additional condition
Eq. (\ref{COM}) so that $S^{\prime}$ determines the
mixing matrix. By doing so, we compare the obtained mixing angle
with the atmospheric neutrino anomaly, which was
reported by the Super-Kamiokande collaboration.
Because the mixing angle is a function of the elements of
$y_{\triangle}$, we can find the region of the elements
which is compatible with the Super-Kamiokande results.
We would like to comment that the origin of the
commutation relation of the Yukawa terms
may be the unification of the quark and lepton.
Because the KM matrix is close to the unit matrix, this may suggest that
up and down quark Yukawa terms may approximately satisfy 
similar commutation relations.  Such a relation may also hold
for the lepton sector in unified theories of quarks and leptons.
The phenomenological study of such systems can be extended to the three flavor
case and may be applied to the solar neutrino problem.
We also note that Smirnov \cite{SMI} studied mixing in the
seesaw model. Because our approach differs from his, and
the transformation to the triangular form is essential
in our approach, the relation between our work and his must
be carefully examined.
We will study these issues in future publications.

\section*{\bf Acknowledgements}

We would like to thank G. C. Branco, H. Minakata, T. Onogi for
discussions and O. Yasuda for valuable comments
on the atmospheric neutrino data. 
The work of T.M. was supported in part by a Grant-in-Aid for
Scientific Research on Priority Areas (Physics of CP violation,
project no.11127210) and by Grand-in-Aid for JSPS Fellows (project no.
98362)  from the Ministry of Education, Science
Culture of Japan.

%%%%%%%%%%%%%%%%%%%%%%%%%%%%%%%
%        REFERENCES           %
%%%%%%%%%%%%%%%%%%%%%%%%%%%%%%%

%%%%%%%%%%%%%%%%%%%%%%%%%%%%%%%
%          FIGURES            %
%%%%%%%%%%%%%%%%%%%%%%%%%%%%%%%
\vspace*{-0.5cm}
\begin{figure}[c]
\begin{center}
\leavevmode
\psfig{file=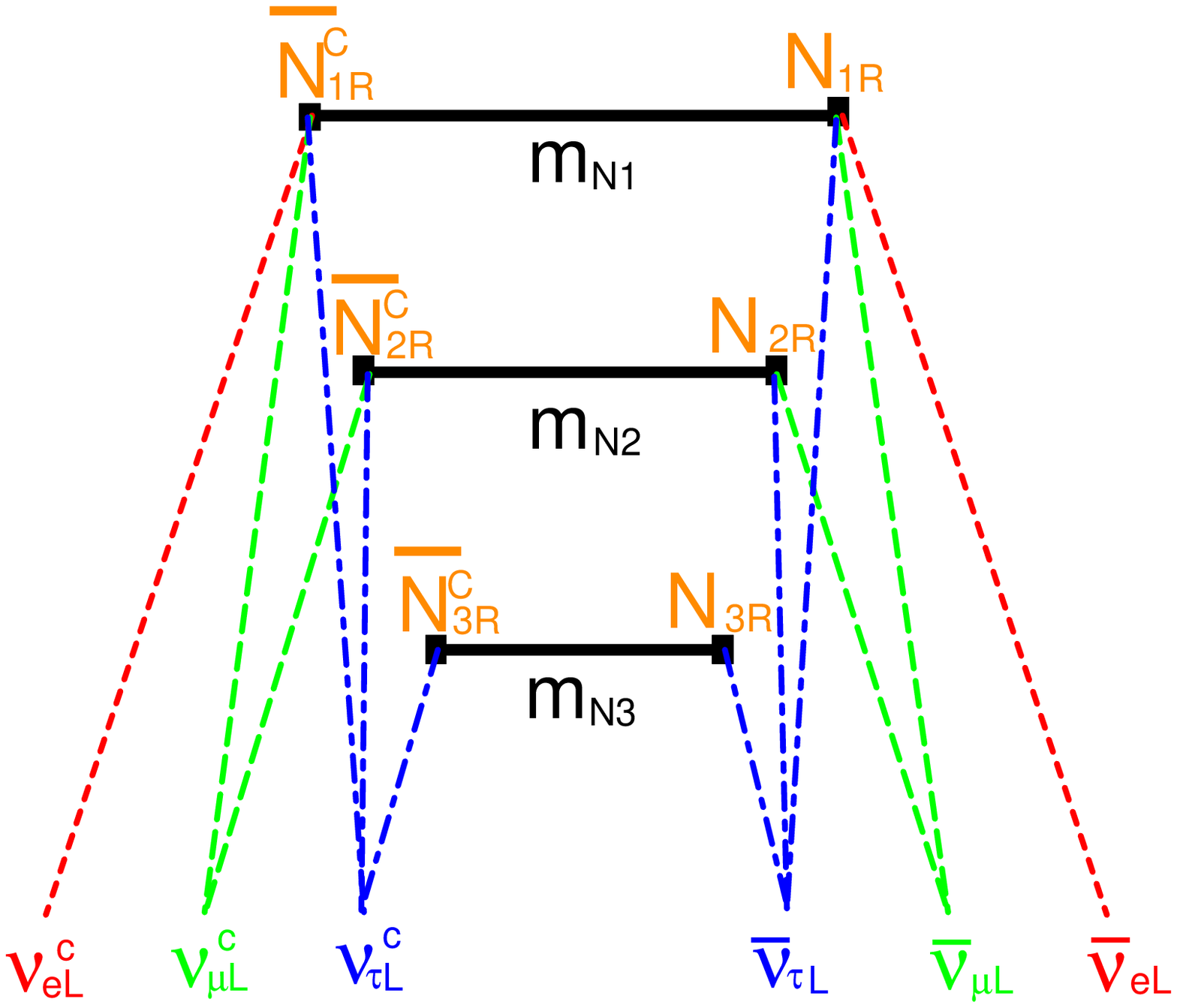,width=10cm}
\caption{In the form of $y_{\triangle}$, $\nu_e$ only
couples to $N_1$, while $\nu_{\mu}$ couples to both $N_1$ and $N_2$,
and $\nu_{\tau}$ couples to all heavy neutrinos, $N_1$, $N_2$ and $N_3$.
The mass of $\nu_e$  depends only on $m_{N1}$.
Moreover, due to the mass hierarchy $m_{N1} \gg m_{N2} \gg m_{N3}$,
$\nu_{\mu}$ and $\nu_{\tau}$ are determined by the masses of
$m_{N2}$ and $m_{N3}$, respectively.}
\end{center}
\end{figure}
%%%%%%%%%%%%%%%%%%%%%%%%%%%%%%
\vspace*{-0.5cm}
\begin{figure}[h]
\begin{center}
\leavevmode
\psfig{file=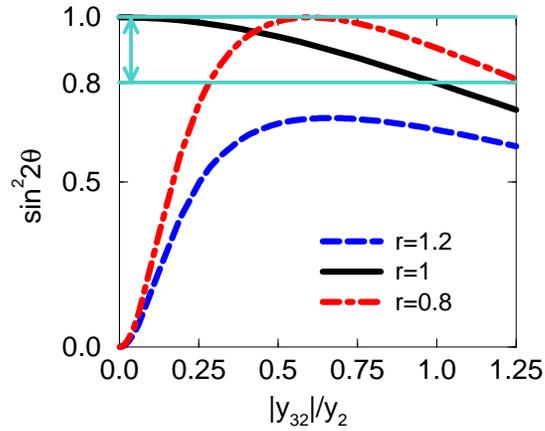,width=8cm}
\caption{$\sin^2 2\theta$ as a function of ${|y_{32}|\over y_2}$
         for various $r={y_3 \over y_2}$. The three curves
         correspond to $r=1$  (solid line), $r=1.2$ (dotted line),
         and $r=0.8$ (dot-dashed line).
         The region between the two solid straight lines is allowed
         by the Super-Kamiokande experiment on 
         atmospheric neutrinos, $0.8 \leq \sin^2 2\theta_{\rm atm} \leq 1$.}
\end{center}
\end{figure}
%%%%%%%%%%%%%%%%%%%%%%%%%%%%%%%
\vspace*{-0.5cm}
\begin{figure}[h]
\begin{center}
\leavevmode
\leavevmode\psfig{file=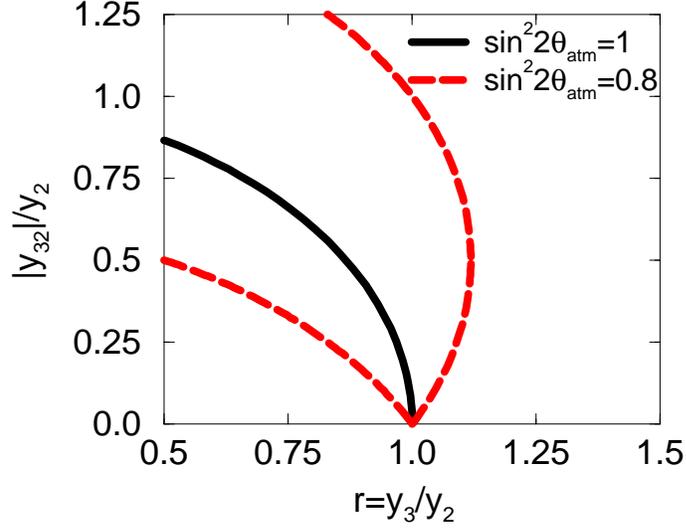,width=10cm}
\caption{The region in the parameter space
         $(\frac{y_3}{y_2}, \frac{|y_{32}|}{y_2})$
         between the dashed lines is allowed
         from the Super-Kamiokande experiment on 
         atmospheric neutrinos, $\sin^2 2\theta_{\rm atm} \geq 0.8$.}
\end{center}
\end{figure}
%%%%%%%%%%%%%%%%%%%%%%%%%%%%%%%
\vspace*{-0.5cm}
\begin{figure}[h]
\begin{center}
\leavevmode
\psfig{file=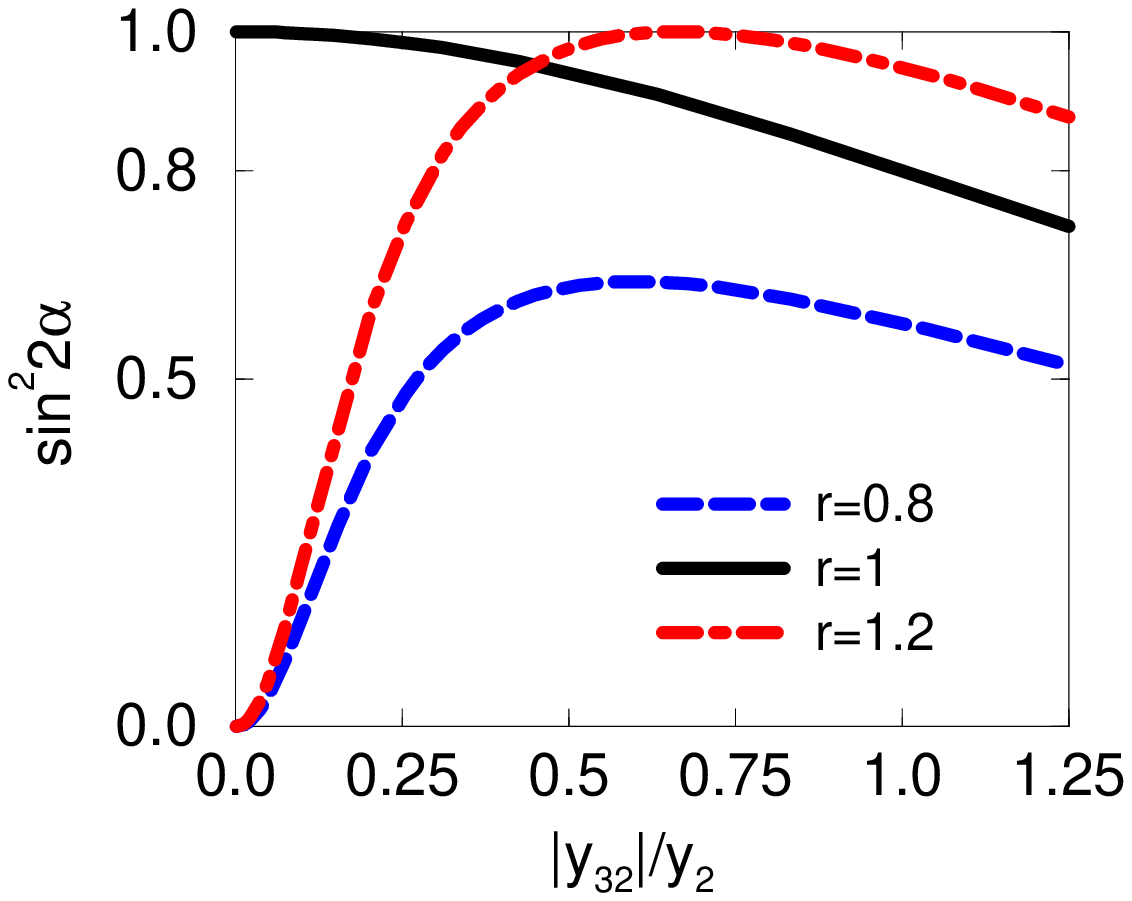,width=8cm}
\caption{$\sin^2 2\alpha$ as a function of ${|y_{32}|\over y_2}$
         for various $r={y_3 \over y_2}$. The three curves
         correspond to $r=1$  (solid line), $r=1.2$ (dotted line),
         and $r=0.8$ (dot-dashed line).}
\end{center}
\end{figure}
%%%%%%%%%%%%%%%%%%%%%%%%%%%%%%%


\baselineskip 18pt

\begin{thebibliography}{1}
%
\bibitem{SOL}
%B. T. Cleveland {\it et al.},
%{\it Nucl. Phys.} {\bf B(Proc. Suppl.) 38} (1995) 47.\\
K.S. Hirata {\it et al.}, {\it Phys. Rev. Lett.} {\bf 65} (1990)
1297; {\it Phys. Rev. Lett.} {\bf 65} (1990) 1301;
{\it Phys. Rev.} {\bf D44} (1991) 2241;
{\bf D45}, (1992) 2170(E).\\
Super-Kamiokande Coll., Y. Fukuda {\it et al.},
{\it Phys. Rev. Lett.} {\bf 77} (1996) 1683;
{\it Phys. Rev. Lett.} {\bf 81} (1998) 1158.\\
J. N. Abdurashitov {\it et al.}, {\it Phys. Lett.} {\bf B328}
(1994) 234.\\
GALLEX Coll., P. Anselmann {\it et al.}, {\it Phys. Lett.}
{\bf B327} (1994) 377; {\it Phys. Lett.} {\bf B342} (1995) 440.
%
\bibitem{ATM}
Y. Fukuda {\it et al.}, {\it Phys. Lett.} {\bf B335} (1994) 237.\\
R. Becker-Szendy {\it et al.}, {\it Nucl. Phys.}
{\bf B(Proc. Suppl.) 38} (1995) 331.\\
W. W. M. Allison {\it et al.}, {\it Phys. Lett.}
{\bf B391} (1997) 491.\\
MACRO Coll., M. Ambrosio {\it et al.}, {\it Phys. Lett.}
{\bf B434} (1998) 451.
%
\bibitem{SKC} Super-Kamiokande Coll., Y. Fukuda {\it et al.},
{\it Phys. Lett.} {\bf B433} (1998) 9;
{\it Phys. Lett.} {\bf B436} (1998) 33;
{\it Phys. Rev. Lett.} {\bf 81} (1998) 1562.
%
\bibitem{MNS} Z. Maki, M. Nakagawa and S. Sakata,
{\it Prog. Theor. Phys.} {\bf 28} (1962) 870.
%
\bibitem{YAN}
T. Yanagida,
in Proceedings of the Workshop on "The Unified
Theory and the Baryon Number of the Universe", Edited by
Osamu Sawada and Akio Sugamoto, KEK 13-14 Feb 1979
(KEK-79-18).\\
M. Gell-Mann,  P. Ramond and R. Slansky in Sanibel Talk,
CALT-68-709, Feb. 1979, and in "Supergravity" (North
Holland, Amsterdam 1979).
%
\bibitem{BIL}
S. M. Bilenky and S.T. Petcov,
{\it Rev. Mod. Phys.} {\bf 59} (1987) 671.
%
\bibitem{LR}
Z. G. Berezhiani, {\it Phys. Lett.} {\bf 129B} (1983) 99;
{\it Phys. Lett.} {\bf 150B} (1985) 177.\\
D. Chang and R. N. Mohapatra, {\it Phys. Rev. Lett.} {\bf 58} (1987) 1600.\\
J. Rajpoot, {\it Phys.Lett.} {\bf B191} (1987) 122.\\
A. Davidson and K. C. Wali, {\it Phys. Rev. Lett.} {\bf 59} (1987) 393.\\
K. S. Babu and R. N. Mohapatra, {\it Phys. Rev. Lett.} {\bf 62} (1989)
1079.\\
Y. Koide and H. Fusaoka, {\it Z. Phys.} {\bf C71} (1996) 459.\\
Y. Koide, {hep-ph/9803458}; {\it Phys. Rev.} {\bf D56} (1997) 2656.
%
\bibitem{MOR}
T. Morozumi, T. Satou, M. N. Rebelo and M. Tanimoto,
{\it Phys. Lett.} {\bf B410} (1997) 233.\\
Y. Kiyo, T. Morozumi, P. Parada, M. N. Rebelo and M. Tanimoto,
{\it Prog. Theor. Phys.} {\bf 101} (1999) 671.
%
\bibitem{BRA}
G. C. Branco and L. Lavoura, {\it Nucl. Phys.} {\bf B278} (1986) 738.\\
G. C. Branco, T. Morozumi, P. A. Parada and M. N. Rebelo,
{\it Phys. Rev.} {\bf D48} (1993) 1167.
%
\bibitem{SMI}
A. Yu. Smirnov, {\it Phys. Rev. } {\bf D48} (1993) 3264.
%
\end{thebibliography}
\end{document}